\ifx \kindle \undefined
\documentclass[]{article}
\usepackage[left=0.75in,top=1.125in,bottom=1.125in,right=0.75in,nohead]{geometry}
\else
\documentclass[fontsize=8pt]{scrartcl}
\usepackage[paperwidth=9cm, paperheight=12cm, textheight=4cm, top=1cm, left=1cm, right=1cm, bottom=1cm]{geometry}
\fi

\author{Stuart B. Heinrich \\ sbheinri@ncsu.edu }
\title{The Relativity of Existence}

\usepackage[numbers,sort]{natbib}

\usepackage[pdftex]{xcolor,graphicx}
\usepackage[linkbordercolor=white,bookmarks=true]{hyperref} %incompatible with float package!
\usepackage{amssymb}
\usepackage[cmex10]{amsmath}
\usepackage{amsthm} %must be loaded AFTER amsmath

\usepackage{verbatim}
\usepackage{url}
\usepackage{cancel}
\usepackage{color}
\usepackage{rotating}
\usepackage{multicol}
\usepackage{boxedminipage}
\usepackage{placeins} % for FloatBarrier
\usepackage{booktabs} % for making nice tables
\usepackage{times} %times font is smaller than default
\usepackage{dialogue}
\usepackage{lips} %for better ellipses than \dots
\usepackage{amsthm}

\newtheorem{theorem}{Theorem}

\newcommand{\secref}[1]{Section \ref{#1}}
\newcommand{\thmref}[1]{Theorem \ref{#1}}

\begin{document}
\maketitle

\begin{abstract}
Despite the success of modern physics in formulating mathematical theories that can predict the outcome of experiments, we have made remarkably little progress towards answering the most fundamental question of: why is there a universe at all, as opposed to nothingness?  In this paper, it is shown that this seemingly mind-boggling question has a simple logical answer if we accept that existence in the universe is nothing more than mathematical existence relative to the axioms of our universe.  This premise is not baseless; it is shown here that there are indeed several independent strong logical arguments for why we should believe that mathematical existence is the only kind of existence.  Moreover, it is shown that, under this premise, the answers to many other puzzling questions about our universe come almost immediately.  Among these questions are: why is the universe apparently fine-tuned to be able to support life? Why are the laws of physics so elegant?  Why do we have three dimensions of space and one of time, with approximate locality and causality at macroscopic scales?  How can the universe be non-local and non-causal at the quantum scale?  How can the laws of quantum mechanics rely on true randomness?
\end{abstract}

\ifx \kindle \undefined
\begin{multicols}{2}
\else
%remove page numbers for kindle
\thispagestyle{empty}
\pagestyle{empty}
\fi

\section{Introduction}

Over the course of modern history, we have seen advances in biology, chemistry, physics and cosmology that have painted an ever-clearer picture of \emph{how} we came to exist in this universe.  However, despite all these advances, it seems we have not made any actual progress towards answering the fundamental question of \emph{why}?

In 510 BCE, Parmenides reasoned that \emph{ex nihilo nihil fit}, or ``nothing comes from nothing,'' meaning that the universe in the now implies an eternal universe without any specific moment of creation.  This viewpoint was shared by later Greek philosophers such as Aristotle and Plato, but does not really answer the question.  In 1697, \citet{LEIBNIZ1697} asked for ``a full reason why there should be any world rather than none.''  He claimed \citep{LEIBNIZ1714} that ``nothing takes place without sufficient reason,'' now known as the Principle of Sufficient Reason (PSR), and generalized the earlier question by asking, ``why is there something, rather than nothing?'', now known as the Primordial Existence Question (PEQ).

The fundamental question, further reviewed in \citet[p.296-301]{Edwards67} and \citet{Lutkehaus99}, has been echoed by many modern philosophers such as Richard Swinburne, who said, ``It remains to me, as to so many who have thought about the matter, a source of extreme puzzlement that there should exist anything at all'' \citep[p.283]{Swinburne91}, and Derek Parfit, who asked, ``Why is there a Universe at all? It might have been true that nothing ever existed; no living beings, no stars, no atoms, not even space or time. When we think about this possibility, it can seem astonishing that anything exists'' \citep[p.24]{Parfit98b}.

\begin{comment}
and, ``We expect all things to have explanations''\citep[p.287]{Swinburne91}, asserting that, ``Surely the most natural state of affairs is simply nothing: no universe, no God, nothing'' \citep[p.48]{Swinburne96}. Derek Parfit has voiced a similar opinion \citep[p.420]{Parfit98a}, asking, ``Why is there a Universe at all? It might have been true that nothing ever existed; no living beings, no stars, no atoms, not even space or time. When we think about this possibility, it can seem astonishing that anything exists'' \citep[p.24]{Parfit98b}.
\end{comment}

Most physicists and cosmologists are equally perplexed.  Richard Dawkins has called it a ``searching question that rightly calls for an explanatory answer'' \citep[p.155]{Dawkins06}, and Sam Harris says that ``any intellectually honest person will admit that he does not know why the universe exists.  Scientists, of course, readily admit their ignorance on this point'' \citep[p.74]{Harris06}.

With that said, modern inflationary cosmology does offer some powerful new insights into this question.  A generic property of inflation is that the universe began from a small quantum fluctuation \citep{Stenger90,Hartle83,Hawking98,Guth07,Guth97book,Pasachoff03} \citep[p.129]{Hawking88}\citep[p.131]{Hawking12}.  According to \citet{Vilenkin83}, ``A small amount of energy was contained in that [initial] curvature, somewhat like the energy stored in a strung bow. This ostensible violation of energy conservation is allowed by the Heisenberg uncertainty principle for sufficiently small time intervals. The bubble then inflated exponentially and the universe grew by many orders of magnitude in a tiny fraction of a second.''

\begin{comment}
Chaotic Inflation theory proposed by Andrei Linde.  According to this scenario, by means of a random quantum fluctuation the universe `tunneled' from pure vacuum (`nothing') to what is called a false vacuum, a region of space that contains no matter or radiation but is not quite `nothing.' The space inside this bubble of false vacuum was curved, or warped. A small amount of energy was contained in that curvature, somewhat like the energy stored in a strung bow. This ostensible violation of energy conservation is allowed by the Heisenberg uncertainty principle for sufficiently small time intervals. The bubble then inflated exponentially and the universe grew by many orders of magnitude in a tiny fraction of a second. \citep{Stenger90}  Modern review: Eternal inflation and its implications \citep{Guth07}

Some physicists have tried to avoid the initial conditions problem by proposing models for an eternally inflating universe with no origin.\citep{Carroll04,Carroll05,Aguirre02,Aguirre03} These models propose that while the universe, on the largest scales, expands exponentially it was, is and always will be, spatially infinite and has existed, and will exist, forever. which may or may not be correct \citep{Unruh98}
\end{comment}

According to Stephen Hawking, ``When one combines the theory of general relativity with quantum theory, the question of what happened before the beginning of the universe is rendered meaningless'' \citep[p.135]{Hawking12}, because, ``when we add the effects of quantum theory to the theory of relativity, in extreme cases warpage can occur to such an extent that time behaves like another dimension of space.  In the early universe--when the universe was small enough to be governed by both general relativity and quantum theory--there were effectively four dimensions of space and none of time'' \citep[p.134]{Hawking12}.

The notion that our timelike dimension emerged out of a spatial dimension is controversial.  It is even more controversial when Hawking argues that, ``The realization that time behaves like space presents a new alternative.  It not only removes the age-old objection to the universe having a beginning, but also means that the beginning of the universe was governed by the laws of science and doesn't need to be set in motion by some God'' \citep[p.135]{Hawking12}, adding, ``Because there is a law like gravity [and quantum physics], the universe can and will create itself from nothing.  Spontaneous creation is the reason there is something rather than nothing, why the universe exists, why we exist'' \citep[p.180]{Hawking12}.  In other words, Hawking believes that Leibniz's question has been answered.

The flaw with this logic is that even if the mathematics of spontaneous creation are correct, they are based on general relativity and quantum physics, which are not ``nothing.''  Thus, it is not the \emph{ex nihilo} creation of something from nothing, but rather, the derivation of the universe in the now from a set of axioms.  This is a trivial result, because for \emph{any} set of non-contradictory statements, one can always derive those statements from a set of axioms by simply taking the statements as axioms.  Hawking has presented an argument that the universe can be derived from a reduced set of axioms, but he has done nothing to answer the question of \emph{why} those axioms (of M-theory, or of general relativity and quantum physics) are true, nor has he shown that they are the most fundamental possible set of axioms.  Thus, it leaves Leibniz' question completely untouched.

Most physicists do recognize this issue.  Brian Greene specifically pointed out that modern inflationary cosmology cannot resolve Leibniz's question \citep[p.310]{Greene04}, adding, ``If logic alone somehow required the universe to exist and be governed by a unique set of laws with unique ingredients, then perhaps we'd have a convincing story.  But to date, that's nothing but a pipe dream'' \citep[p.310]{Greene04}.

The fundamental problem that has prevented Leibniz's question from being answered is the blind faith assumption that the universe really does exist in an absolute, objective sense.  When one releases this faith, and begins to consider the possibility that there are multiple independent universes, which do not share any common laws of physics (i.e., not simply embedded in a common framework or multiverse), then answers come much more freely.

Under these lines, a theory that very nearly meets Greene's goal has been proposed by \citet{Tegmark98}, known as the Mathematical Universe Hypothesis (MUH) or Ultimate Ensemble theory \citep{Tegmark03,Tegmark08}.  As formulated by Tegmark, the MUH rests on the sole postulate that ``all structures that exist mathematically also exist physically'' \citep{Tegmark98}.  Unfortunately, the meaning of this initial postulate is difficult to grasp and apparently lacks reasonable justification.

The primary purpose of this paper is to show that Tegmark's hypothesis, which would be better stated as ``physical existence is the same thing as mathematical existence'' is really just an implication of the Relativity of Existence (ROE), and that the ROE can be logically derived independently in several different ways, such as from the postulate that human self-awareness is governed by the laws of physics (\secref{sec_axiomatization}), or from a logical paradox proving that ``if we exist in any sense, then that existence is relative rather than truly objective'' (\secref{sec_infinite_regress}).

\section{Logical Arguments} \label{sec_logic}

This paper presents four independent logical arguments in support of the ROE.  The first argument (\secref{sec_thought_experiment}) is a humorous thought experiment that intuitively validates the ROE, but is not a logically sound argument.  The second argument (\secref{sec_axiomatization}) is a direct logical derivation of the ROE from the sole assumption that self-awareness can be represented with finite information.  The third argument (\secref{sec_infinite_regress}) shows that the concept of a single objective reality is an oxymoron, and that if we exist, our existence must be relative.  Finally, the fourth argument (\secref{sec_anthropic}) shows that the ROE is the only possible theory that could ever fully answer Leibniz's question or resolve the fine-tuning problem.

\subsection{The Simulation Thought Experiment} \label{sec_thought_experiment}

If self-awareness evolved in humans by emergent processes that are subject to the laws of physics, and the laws of physics are consistent mathematical (or statistical) rules, then there seems to be no fundamental reason why these consistent mathematical rules could not be programmed into a computer in order to simulate a virtual world, with equal or similar laws of physics, in which self-aware life could also evolve.  

This is an old concept, and some have even gone so far as to suggest that our universe is really just a simulation (e.g., see \citet{Drexler85,Tripler94,Schmidhuber97,Bostrom98,Moravec99,Schmidhuber00,Bostrom03,Kurzweil99}).  To be clear, the idea that our universe is really just a computer simulation is highly controversial \citep{McCabe05}, and not supported by this author.  However, let us entertain the idea that someday in the future, the computational challenges are overcome and the laws of physics are understood well enough that a simulation containing self-aware life could actually be created.  

Suppose that, in this future, a researcher named Bob does actually create a virtual world and observes the evolution of self-aware life forms, which look something like a primitive colony of humans.  Through the window of his simulation, Bob observes them being born, dying, hunting for food, falling in love, etc.  In short, they are observed to experience every emotion that we attribute to humanity.

Would these simulated beings be aware of the processing time that it takes Bob to simulate outcomes in the virtual world?  The answer, as most would agree, is clearly not -- the perceptions of these virtual beings is limited to the axioms of their world, and the processing time only affects how rapidly Bob can simulate their world without actually affecting the internal dynamics of that world.  Likewise, if Bob paused the simulation to go out to lunch and then returned to the laboratory to resume the simulation, most agree that the simulated beings would not detect any change.

What would happen if Bob's funding was cut, causing him to abandon his research and turn off the simulation never to be started again?  Does this affect the self-aware beings?  Does their world suddenly cease to exist?  If the reader is inclined to say that their world ceases to exist, then consider what happens when, 10 years later, a graduate student named Eric manages to recover the lost hard drives and restart the simulation.  Because this is nothing but a long pause, we should expect the life forms to pick up life exactly where they left off without any awareness of the external pause.

Suppose that Eric has a mean streak and wants to play God by programming in a mysterious hand that appears out of nowhere, picks up the poor little virtual people, and squishes their heads in.  Has he actually affected the virtual world?  On the one hand, yes -- because he can see their loved ones crying over the loss, and he can see their philosophers tearing their hair out over the sudden violation of the physics they thought they understood so well.  On the other hand, Bob might learn of Eric's mischief and decide to roll-back the simulation and proceed from the time period before Eric's interventions, thereby nullifying those effects.

Of course, nothing is stopping Eric from making a copy of the simulation and continuing to observe what happens while he plays God.  Thus, the two researchers could observe parallel worlds unfolding before their eyes.  It is not difficult to imagine that two different parallel universes could be created in this manner, because if one universe can be created in a simulation, then so can two.  Indeed, one realizes that there are an infinite number of universes that could be created by changing any event at any point in time, and Eric's intervention merely changed which one of those infinite number of universes is being observed through the simulation.

Clearly, the simulation is necessary in order to observe what happens in the virtual world, but does the act of running the simulation make the world become real from the internal perspective of those self-aware beings?  Clearly, the physical machine does not matter -- the simulation could be transferred to a different machine, or even continued on pencil and paper (albeit slowly), or even perhaps in the mind of someone who observed the state of the computer's registers and was able to mentally predict what happens next.  What difference does it make from the internal perspective of a self-aware being if the constructs of their world are simulated or imagined from an external world?

Under the Mathematical Universe Hypothesis (MUH), the simulation does not need to be run in order to make a potential universe real, because all possible worlds exist in a timeless mathematical sense without the need for human derivation or observation \citep{Tegmark08}.  However, it would be would be wrong to say that these realities exist in `parallel' because that implies a unifying fabric of time between them.  A simulation can be started, stopped, and restarted to observe the exact same behaviors, so there is clearly no shared fabric of time.

Can this thought experiment be taken as logical evidence for the MUH?  This question is investigated in further detail in the next section, where the assumptions and conclusions of the thought experiment are made explicit by way of a more well defined logical argument.

\subsection{The Axiomatization of Self-Awareness} \label{sec_axiomatization}

By the principle of explosion, in any system that contains a single contradiction, it becomes possible to prove the truth of any other statement no matter how nonsensical\citep[p.18]{Franzen05}.  There is clearly a distinction between truth and falsehood in our reality, which means that the principle of explosion does not apply to our reality.  In other words, we can be certain that our reality is \emph{consistent}.

Any system with finite information content that is consistent can be formalized into an axiomatic system, for example by using one axiom to assert the truth of each independent piece of information.  Thus, assuming that our reality has finite information content, there must be an axiomatic system that is isomorphic to our reality, where every true thing about reality can be proved as a theorem from the axioms of that system.

Conscious self-aware life forms (such as humans) exist in our reality.  Thus, it must somehow be possible to derive the existence of each such being in our reality as a theorem from the axioms of our reality.  This logic applies to all aspects of those beings, including the actual experience of consciousness or self-awareness.

Because an axiomatic system is defined by a unique set of axioms, for any axiomatic system that can derive a particular theorem, there is another axiomatic system that also derives that theorem.  This other system can be found by modifying the original set of axioms in such a way that the theorem can still be derived.  For example, by the simple inclusion of a new axiom that does not contradict any existing axioms, or by modifying or subtracting an axiom that was not directly used in the derivation of the theorem.\footnote{As a concrete example, suppose that the popular Copenhagen interpretation of quantum mechanics is true.  In this model, the laws of physics describe only the probability of obtaining a measurement, meaning that the specific outcome of a measurement (or more specifically, the collapse of a wavefunction) requires a unique one-shot axiom to describe that outcome in the axiomatic system of our reality.  Because no particular outcome is a violation of the laws of physics, \emph{any} outcome for an individual measurement is plausible; hence, if we consider the axiomatic system formed by changing the outcome of a single quantum measurement, it will necessarily be a different system that is in some ways contradictory to our reality, yet still capable of deriving the presence of self-aware life.}  Clearly, there are an infinite number of ways to modify an axiomatic system while keeping any particular theorem intact.

Thus, there must be an infinite number of other axiomatic systems that are in some ways different from our reality, yet still can derive self-aware beings with self-aware thoughts.  But are these other axiomatic systems real, or merely hypothetical?  From an objective perspective, none of these systems are real -- they are all based on sets of axioms that contract each other.  However, from the perspective of a self-aware being within any one system, everything that is observed appears completely consistent and real.

I have no basis for believing that my universe is real other than the fact that I observe a consistent world around me, but the same thoughts can be proven to be taking place in other axiomatic systems that are contradictory to the system that describes my reality!  In other words, my reality is no more real than the others.  There is no way to objectively identify any one axiomatic system as being 'the real one.'  None of them are (objectively) real, and at the same time, all of them are (subjectively) real.  This is the relativity of existence.

Recognizing this, the ultimate answer to the question of why our reality exists becomes trivial: it doesn't exist at all, not objectively.  However, somehow it is possible for self-awareness to be represented axiomatically, and this means that any axiomatic system that can derive self-awareness \emph{is} perceived as being real (internally) without the need for an objective manifestation.  Thus, there is nothing special about our reality, it is just one of an infinite number of completely disjoint mathematical spaces.

\subsection{The Relativity of Existence} \label{sec_infinite_regress}

As noted by Greene, ``Even if a cosmological theory were to make headway on this [Leibniz's] question, we could ask why that particular theory--its assumptions, ingredients, and equations--was relevant, thus merely pushing the question of origin one step further back'' \citep[p.310]{Greene04}.  This problem of infinite regress has been well understood since antiquity \citep[p.38]{Franzen05}.  In fact, it is more than just a curious question, because our natural assumptions on this subject lead to a logical contradiction.

If a sentence S is objectively true, then ``not S'' is a contradiction, and a contradiction among some finite system of axioms is provably false.  However, a proof that ``not S'' is false is also a proof that ``S is true,'' which means that any \emph{objectively} true statement can be proven true with an \emph{objective} proof.

Not many things can be proven objectively true, because any proof relying on axioms is not objective without proving that the axioms are also objectively true.  A proof requires a finite sequence of steps, so it cannot be an infinite regress.  Thus, in order to be truly objective, a proof must be free of all axioms.  However, the only things that can be proven true without axioms are simple tautologies.

\begin{theorem} \label{thm_objective_truth}
The only things that are objectively true are tautologies.
\end{theorem}  

Clearly, our existence is not a tautology.  Presumably, our existence can be derived from the axioms of our reality (i.e., the general laws of physics, along with any nuance axioms to describe initial conditions or the outcome of random events), but these axioms are not tautologies.  Thus, by \thmref{thm_objective_truth}, it cannot be objectively true that we exist.

The only resolution to this paradox is to recognize that the assumption that our existence is \emph{objectively} true is an oxymoron that must be rejected.  This doesn't mean that Ren\'{e} Descartes' famous conclusion of \emph{cogito ergo sum}, or ``I think, therefore I exist,'' was \emph{wrong}; rather, it means that existence is not objective, but relative to some set of axioms, and that \emph{relative existence does not preclude the self-aware experience}.  Thus, our preconceived notion of physical existence is no different from mathematical existence \citep{Hilbert34} relative to the arbitrary set of axioms that define our reality.

\begin{theorem}
The self-aware experience can be derived mathematically from some set of axioms.  From the perspective of this self-awareness, mathematical existence is physical existence.
\end{theorem}

\subsection{The Anthropic Principle} \label{sec_anthropic}

Habitability of a planet depends on a confluence of factors ranging from parent star class \citep{Kasting93,Kasting97} and stellar variation \citep{Lammer07}, to planet mass \citep{Raymond07}, composition, orbit distance \citep{Huggett95}, stability \citep{Lasker93}, early geochemistry conditions \citep{Parkinson08} and many other factors \citep{Irwin01}.  If all of these properties were chosen at random, without any overall guiding influence or purpose, then the statistical likelihood of achieving conditions amenable to life must be exceedingly small.  Moreover, even on a theoretically habitable world, when we consider the likelihood of random chemistry interactions giving rise to self-replication and the actual evolution of life, our assessment of the overall likelihood of any random planet harboring life becomes even more diminutive.

Without knowing the precise details of all the chemical interactions that took place in order to give rise to life on Earth, it is difficult to make accurate predictions as to how small this likelihood actually is.  However, despite recent observations of potentially habitable exoplanets \citep{Borucki11} and better models of early biochemistry indicating that life might not be quite as rare as originally believed \citep[p.47]{Kauffman95}, it is only by taking into account our cosmological observations of billions upon billions of other star systems that we can explain the presence of life as something to be truly expected.  By reasoning of the anthropic principle that ``conditions observed must allow the observer to exist,'' it is truly only necessary for at least one of the practically infinite number of planets in the universe to contain life in order for us to resolve the mystery of why, when we look around, we should observe a planet with all the right conditions for life \citep{Penrose91}.

However, the mystery is still not fully solved, because it merely illustrates the remarkable perfection of the underlying laws of physics that gave rise to a universe containing the capacity for life.  From the molecular properties of water \citep{Henderson1913} to the precise balance of forces such as gravity and electromagnetism \citep{Dicke61}, to the number of dimensions and the precise values of all the fundamental constants, all of which exist in a perfect balance.

As stated by Paul Davies, ``There is now broad agreement among physicists and cosmologists that the universe is in several respects ‘fine-tuned' for life \citep{Davies03}.''  According to Stephen Hawking, ``The laws of science, as we know them at present, contain many fundamental numbers, like the size of the electric charge of the electron and the ratio of the masses of the proton and the electron...and the remarkable fact is that the values of these numbers seem to have been very finely adjusted to make possible the development of life'' \citep[p.125]{Hawking88}.  For example, if the strength of the strong nuclear force were changed by a mere 2\%, the physics of stars would be drastically altered so much that all the universe's hydrogen would have been consumed during the first few minutes after the big bang \citep[p.70-71]{Davies93}.

Can we again invoke the anthropic principle in order to answer this question, once and for all?  This second application is known as the ``strong'' anthropic princple (SAP) \citep{Barrow88}.  However, a second application requires some evidence that there are an extremely large number, perhaps infinite, of different universes having different physical laws.

It is believed by some that the modern incarnation of superstring theory known as M-theory \citep{Duff96} satisfies this condition.  Under M-theory, there are 11 dimensions of spacetime, 7 of which have been curled up into some Calabi-Yau manifold \citep{Candelas85}, and the fundamental constants can be derived from the way that the dimensions have been curled up\citep[p.372]{Greene04}.  Because there are at least $10^{500}$ different ways to curl up these dimensions \citep[p.118]{Hawking12}, and the theory does not dictate which way is correct, it is believed that all ways are equally valid and that the selective power of the SAP explains why we exist in a universe with fundamental constants amenable to life.  The different configurations are interpreted either as parallel universes within the multiverse \citep[p.93]{Kaku05}, or as parallel histories of the same universe \citep[p.136]{Hawking12}.

The problem with these explanations based on M-theory is that they answer the fine-tuning problem only partially.  Even if all the configurations allowed by M-theory were manifested, this would not explain why the underlying mathematics of M-theory were true.  One could just as easily ask why the axioms of M-theory had been miraculously selected in order to give rise to a multiverse capable of supporting life.

Only a theory that admits all possible axiomatic systems can fully resolve the fine-tuning problem, because only when we stop idolizing the axioms of our universe as being objectively special can we stop questioning why they are special.  This is precisely what is offered by the ROE: it tells us that there is nothing special about our universe, because any axiomatic system that can derive self-awareness will be real from the perspective of that self-awareness.

\section{Refutation of Common Objections} \label{sec_refutations}

A number of overall objections to the MUH have already been summarized and refuted by \citet{Tegmark08}.  Therefore, this section will mostly focus on refuting particular objections to the logical arguments for the ROE presented in \secref{sec_logic}; most notably, Tegmark's own misgivings about G\"{o}del's theorems (\secref{sec_incomplete}).

\subsection{Limitations of Anthropic Reasoning} \label{sec_anthropic_limits}

It has been claimed that ``anthropic reasoning fails to distinguish between minimally biophilic universes, in which life is permitted but only marginally possible, and optimally biophilic universes, in which life flourishes because biogenesis occurs frequently''\citep{Davies03}.  This point is misguided, because it implies that we should be compelled to explain every aspect of our universe with high likelihood.  If a theory predicts all the conditions of our universe with high likelihood, then that is indeed encouraging, but it should not be troubling to us if certain aspects of our universe are regarded as having a `low' or less than maximal likelihood, as long as such universes are allowed to exist under the theory.

Regardless, anthropic reasoning does indeed succeed at selecting optimally biophilic universes.  Consider the simple case of two universes, one being a minimally biophilic universe harboring 1 self-aware life form, and the other being an ``optimally'' biophilic universe harboring 1 million self-aware life forms.  In this simple example, any self-aware being selected at random will have a 99.9999\% chance of observing its surroundings to be an optimally biophilic universe.  This logic continues to hold as the number of universes approaches infinity.

Moreover, any axiomatic system defining some simple and elegant laws that can give rise to the evolution of life as an emergent process will almost assuredly be teeming with life, because those emergent processes will repeat themselves over and over.  Thus, the anthropic principle will indirectly select for universes with simple and elegant laws as well, so we should not be surprised that our universe has such a set of simple and elegant laws.

One realizes that it is impossible to have emergent processes without at least an approximate notion of causality, because without causality there can be no change.  It is also impossible to have emergent processes without at least some approximate notion of locality, because without spatial relationships there could be no shape, form, structure or complexity in the universe.  Thus, the anthropic principle also selects for axiomatic systems with spacelike and timelike dimensions, so we should not be surprised to observe those.

Finally, we should not be surprised to find that, at the smallest quantum scale, the universe is not \emph{perfectly} local or causal, because there is no difficulty in representing non-localities or temporal dependencies in axiomatic systems, and anthropic reasoning can only select for local and causal properties insofar as they permit the macroscopic capacity for emergent processes.

\subsection{G\"{o}del's Theorems} \label{sec_incomplete}

Formally, an axiomatic system is called \emph{consistent} if it cannot prove any statement along with its negation (a contradiction), and \emph{complete} if every sentence that can be expressed in the language can be either proved or disproved.  G\"{o}del's first theorem shows that any axiomatic system containing a modicum of arithmetic power is incomplete, and his second theorem shows that any axiomatic system containing a modicum of arithmetic power cannot prove its own consistency \citep{GodelWorks86}.  These theorems have been the subject of many confusions and misunderstandings, as summarized in \citet{Franzen05}.

Currently, all theories of physics are very mathematical, and hence both of G\"{o}del's theorems apply.  With regards to the first theorem, there is a commonly expressed fear that no theory of physics will ever be able to fully describe all aspects of reality, because ``there will always be some truths about the real universe that cannot be proven.''

For example, in his 2003 lecture at the Cambridge-MIT Institute (CMI), Stephen Hawking said, ``According to the positivist philosophy of science, a physical theory is a mathematical model.  So if there are mathematical results that cannot be proved, there are physical problems that cannot be predicted,'' adding, ``...some people will be very disappointed if there is not an ultimate theory, that can be formulated as a finite number of principles. I used to belong to that camp, but I have changed my mind'' \citep{Hawking03lecture}.

This sentiment was echoed by Freeman Dyson, who said, ``His theorem implies that pure mathematics is inexhaustible. No matter how many problems we solve, there will always be other problems that cannot be solved within the existing rules. Now I claim that because of G\"{o}del's theorem, physics is inexhaustible too'' \citep[p.225]{Freeman06}, and also Mark Alford, who said ``The methods allowed by formalists cannot prove all the theorems in a sufficiently powerful system [because of G\"{o}del's theorem].  The idea that math is `out there' is incompatible with the idea that it consists of formal systems'' \citep{Hut06}.

However, as pointed out in a response by Solomon Freeman and later conceded by Dyson, ``The basic equations of physics, whatever they may be, cannot indeed decide every arithmetical statement, but whether or not they are a complete description of the physical world, and what completeness might mean in such a case, is not something that the incompleteness theorem tells us anything about''\citep[p.88]{Franzen05}.

In other words, the first incompleteness theorem does \emph{not} imply that there will always be some truths that cannot be proven \citep[p.24]{Franzen05}.  Indeed, there is no restriction against having an `incomplete' axiomatic system where every \emph{provably true} theorem corresponds to a true statement about reality, and every true statement about reality also corresponds to a theorem.  If the ROE is correct, then reality is defined by the things that are provably true, and any additional undecidable statements simply have no bearing on that reality.

Tegmark has also expressed doubts with regards to the second theorem, lamenting that, ``Our standard model of physics includes everyday mathematical structures such as the integers (defined by the Peano axioms) and real numbers. Yet G\"{o}del’s second incompleteness theorem implies that we can never be 100\% sure that this everyday mathematics is consistent: it leaves open the possibility that a finite length proof exists within number theory itself demonstrating that 0 = 1.  Using this result, every other well-defined statement in the formal system could in turn be proven to be true and mathematics as we know it would collapse like a house of cards'' \citep[p.21]{Tegmark08}.

In order to escape this issue, Tegmark proposed the more restricted Compute Universe Hypothesis (CUH) \citep{Tegmark08} as an alternative to the MUH, which only includes axiomatic systems that are simple enough to escape these G\"{o}del-inspired worries.  However, Tegmark's fears in regard to the second theorem are also unfounded.

As explained by \citet[p.101]{Franzen05}, ``The second incompleteness theorem is a theorem about \emph{formal provability}, showing that...a \emph{consistent} theory T cannot postulate \emph{its own consistency}, although the consistency of T can be postulated in another consistent theory,'' adding, ``...it does not tell us whether `T is consistent' can be proved in the sense of being shown to be true by a conclusive argument, or by an argument acceptable to mathematicians.''

In other words, this is essentially the same as the Halting problem \citep[p.173]{Sipser06}, where finding a contradiction is akin to halting the program.  If the program runs indefinitely, or if one searches indefinitely for a contradiction without finding one, then the halting problem or the consistency-check cannot be completed.  However, the ROE as derived in this paper does not differentiate between those formal systems that can \emph{prove} their own consistency or not; simply being consistent is all that matters.  Thus, the only real implication of the second incompleteness theorem is that humans can never fully \emph{verify} that our consistent theories are consistent, or that our attempts to determine the exact axioms of reality are correct.  This is not a new concept, it's the reason why scientists use the word `theory' rather than `fact.'

\subsection{Irrelevance of Leibniz's Question}

According to the logic of \secref{sec_infinite_regress} and \secref{sec_anthropic}, the ROE is the only theory that can answer Leibniz's question.  However, some philosophers, such as \citet{Bergson74}, have criticized Leibniz's question on the grounds that, ``the question presupposes that reality fills a void, that underneath Being lies nothingness, that \emph{de jure} there should be nothing, that we must therefore explain why there is \emph{de facto} something.'' 

\citet[p.5]{Grunbaum00} agrees, essentially arguing that the only reason for Leibniz to ask ``why is there something, rather than nothing?'' is because of the parsimonious principle that ``simple theories are objectively more likely to be true than are complex ones'' \citep{Smart84}, which was incorrectly interpreted to mean that the null theory, which derives nothing, should be infinitely likely.  He calls this argument the ontological Spontaneity of Nothingness (SoN), argues that it is flawed, and that Leibniz's question is thereby rendered meaningless.

The SoN is indeed false, as can be recognized by considering the logical basis of parsimony.  Specifically, a `theory' is any logical argument that can derive some observation as a logical consequence from a set of axioms.  If all the axioms are regarded as being independent and equally likely, then the overall likelihood of the theory is merely the product of the likelihood of its axioms.  Thus, if simplicity is regarded as the number of axioms, then a simpler theory is objectively more likely.

However, because this principle only applies when the axioms can be meaningfully assigned likelihoods, one would need to make the implicit assumption that the axioms describing the physical laws of the universe could be assigned likelihoods in order to apply parsimony to a theory explaining the origin of the universe.  If this were truly the case, it would imply the existence of an even more fundamental set of axioms that describe a statistical framework from which universes are created with random physical laws.  That would mean the initially assumed axioms are actually theorems, not axioms.  In other words, the concept of ascribing likelihoods to the axioms of reality is an oxymoron.  Thus, the axioms of our reality cannot be probabilistic, and hence one cannot say that a world defined by a simpler (or null) set of axioms is ``more likely.''

However, it is not the SoN that makes Leibniz's question relevant, so discrediting the SoN does not make the question meaningless.  Rather, Leibniz's question is relevant because as shown in \secref{sec_infinite_regress}, objective truths must be objectively provable without axioms, yet our existence is far too complex to be objectively proven without axioms.  Under the assumption that our existence is objective, this would seem to disprove our existence.  In the face of this paradox, the question of ``why do we exist?'' is indeed worthy of being asked.

\subsection{Infinite Information} \label{sec_infinite}

One potential argument against the logic of \secref{sec_axiomatization} is that the universe has an infinite information content, thereby preventing it from being represented by an axiomatic system.

Most cosmologists believe that there is a finite amount of energy in the observable universe, with recent analysis of 7-year data from the Wilkinson Microwave Anisotropy Probe (WMAP) estimating that of this finite amount, $72.8\% \pm -1.6\%$ is in the form of Dark Energy, $22.7\% \pm 1.4\%$ is in the form of Dark Matter, and $4.56\% \pm 0.16\%$ is in the form of regular baryonic matter \citep{Jarosik11,Komatsu11,Gold11,Larson11,Bennet11,Weiland11}.

However, this is just the observable universe, and we may still wonder if the universe has infinite extent.  According to the Friedmann-Lema\^{i}tre-Robertson-Walker (FLRW) model (or Standard model of cosmology \citep{Bergstrom06}), there are three possible overall `shapes' of the universe described by the curvature, $\Omega_k$, which can be deduced based on the density of matter.  

If $\Omega_{k} = 0$ exactly then the universe is flat and infinite, if $\Omega_k > 0$ then the universe is spherical and finite (and the curvature also tells us the size \citep{Milnor82}), and if $\Omega_k < 0$ then it is hyperbolic and infinite.  So far, the data has shown that the curvature is very \emph{close} to 0 \citep{Komatsu11}, but this is expected, and is insufficient to determine the sign \citep{Lesgourgues08}.  Indeed, if the magnitude of the true curvature is less than $10^{-4}$, then it might never be possible to determine by any future experiment \citep{Vardanyan09}.

Regardless, it is widely believed that the total positive energy of matter is exactly canceled out by the negative energy of gravity, thereby allowing the entire universe to be created out of the small amount of uncertainty in the vacuum energy of free space \citep[p.129]{Hawking88} \citep{Hartle83,Hawking98,Guth07,Guth97book,Pasachoff03} \citep[p.180]{Hawking12}.  If this is true, then the amount of positive energy is necessarily finite.

In addition, the Bekenstein bound \citep{Bekenstein08}, which can be derived from consistency between the laws of thermodynamics and general relativity \citep{Bekenstein81,Jacobson95,Bousso99,Bousso99b, Bekenstein00, Bousso02,Bousso03, Bekenstein05,Bekenstein08}, tells us that there is a finite information content in any finite region of space containing finite energy.  

Thus, even if the universe did have infinite spatial extent and infinite energy (contrary to modern inflationary cosmology), there would still be a finite information content to any particular region of the universe (such as a galaxy).  This means that at the very minimum, an arbitrarily large finite region can be represented by an axiomatic system, and assuming that self-aware life can evolve in a finite region of space, then the logic of \secref{sec_axiomatization} would still hold in that region.

\subsection{Impossibility of Quantum Simulation}

It has been suggested that self-awareness (i.e., consciousness) might rely on quantum effects \citep{Jibu95,Penrose91,Vitiello01,Ritz04,Conte07,Conte09}, although this remains a minority opinion \citep{Searle97,Chalmers03}, and it has been argued that quantum decoherence disproves the quantum mind hypothesis \citep{Seife00,Tegmark00mind}.  Still, a supporter of the quantum mind theory might potentially argue that quantum effects are fundamentally impossible to simulate, thereby rendering the thought experiment of \secref{sec_thought_experiment} meaningless.  That argument is investigated here, although the reader is first reminded that \secref{sec_thought_experiment} was only intended as a thought-primer, and is independent from the logical arguments of \secref{sec_axiomatization}, \secref{sec_infinite_regress} and \secref{sec_anthropic}.

As summarized by \citet{Chalmers95}, ``It is natural to speculate that these [quantum] properties may play some role in the explanation of cognitive functions, such as random choice and the integration of information, and this hypothesis cannot be ruled out a priori. But when it comes to the explanation of experience, quantum processes are in the same boat as any other. The question of why these processes should give rise to experience is entirely unanswered.''

Indeed, given that there is now good evidence that quantum coherence is involved in photosynthesis \citep{Sarovar10,Panitchayangkoon11}, it is not difficult to imagine that quantum effects are somehow involved in brain function.  However, even if this is the case, there is still no logical reason why the overall effects could not be approximated by some similar method in a computer simulation.  Moreover, there is no property of quantum physics that we know of that could explain the sensation of experience any better than classical physics.

The potential difficulty in simulating quantum systems comes from the original Copenhagen interpretation, where the belief is that particles do not have definite existence or properties while they are not being observed.  The theory states that inbetween periods of observation, the probability of an experimenter obtaining a measurement is described by a probability wavefunction, which collapses into a definite value upon observation.  If this theory is the \emph{fullest} description of reality, as is believed by its constituents, then simulation is impossible because there simply is no mathematics or logic that can be used to describe the interactions within a system while it is not being observed, and there is no way to observe a system without eliminating quantum effects.

\begin{comment}
It is worth noting that, while statistical predictions made under the Copenhagen interpretation have proven remarkably accurate, it has never been demonstrated via simulation that this mathematics can even describe the molecular stability of a single molecule, because the conceptual interpretation of the theory simply does not permit simulation.
\end{comment}

It was originally thought that the Einstein-Podolsky-Rosen (EPR) paradox \citep{Einstein35}, Bell's inequality \citep{Bell64}, and Aspect's experiments \citep{Aspect82a,Aspect82b} were proof that the Copenhagen interpretation must be correct.  However, it is now widely understood that these experiments merely prove that non-local interactions are possible via quantum entanglement \citep[p.114]{Greene04}.  The modern physics community is now divided between several different interpretations of quantum mechanics \citep[p.208]{Greene04}, including the Copenhagen interpretation, Ghirardi-Rimini-Weber theory \citep{Ghirardi86}, the Many Worlds interpretation \citep{Vaidman02online}, and Bohmian mechanics \citep{Pladevall12}.  Each interpretation has slightly different mathematics that lead to slightly different physical predictions, so the differences are not merely philosophical.

All of the other interpretations do permit simulation, so until the Copenhagen interpretation is the last one standing, it would be overly hasty to assume that quantum physics cannot be simulated.

\begin{comment}
In contrast, GRW Flash ontology, Bohmian mechanics, and the Many Worlds interpretation all permit simulation during periods of non-observation.  GRW Flash ontology maintains the statistical nature of the wavefunction but postulates that collapse of the wavefunction can be induced by interactions with other particles in close proximity rather than observers, thereby giving way to predictable Newtonian effects of large structures and gracefully degrading to more unpredictable quantum effects for isolated particles.  Bohmian mechanics is fully deterministic, and theorizes that the wavefunction is a separate medium that permeates space and acts as a guiding force for particles.  Finally, under the Many Worlds interpretation, every possible random outcome corresponds to a different universe, and our probabilistic assessments about measurements are really just quantifications of the probability of being in a universe where that measurement has a particular outcome.  The Many Worlds view is an interpretation that is independent of the mathematics, and could be combined with, say, GRW Flash interpretation.
\end{comment}

\subsection{Quantum Randomness}

Tegmark has claimed that the MUH is incompatible with true quantum randomness because it is impossible to generate a sequence of true random numbers using only axiomatic relationships \citep[p.10]{Tegmark08}.  While it is true that random numbers cannot be generated algorithmically, this does not mean that the physics of our universe must be describable in a deterministic way.  In fact, even before the presence of infinitely many worlds were predicted bottom-up by the MUH, they had been theorized about as a top-down explanation for quantum randomness under the Many Worlds interpretation of quantum physics.

In order to see how this is possible, consider a simple universe containing a particle with position $X$ parameterized by time $t$, with an axiom that simply states,

\begin{equation}
|| X(t) - X(t+1) || \leq \tau \text{.}
\end{equation}

This constraint sets up a random walk, but there are clearly an infinite number of different ways to choose the sequence $X(t)$ that comply with this constraint, all corresponding to different axiomatic systems.  From the internal perspective of an observer within any one axiomatic system, $X(t=now)$ can be measured, but $X(t-1)$ and $X(t+1)$ can only be guessed based on the general laws of physics that could be inferred from past measurements.  In this case, the only general axiom that could be inferred from past measurements is the constraint that set up the random walk.  Thus, if an internal observer were to formulate the laws of physics, they might say,

\begin{quote}
 ``If a particle is observed at $X(t)$, then $X(t+1)$ will be uniformly randomly distributed in the range $( X(t)-\tau, X(t)+\tau )$.''
\end{quote}

Thus, even though the true axioms of this system had no notion of random numbers, the best possible laws of physics formulated by an internal observer does.  This is because the probability represents the observer's uncertainty in being able to predict all the ``nuance'' axioms that specify the precise values of $X(t)$ for $t = 0,1,2 \ldots$.

At present, the Many Worlds interpretation does not consist of a formal set of non-random axioms that can predict the exact shape of the distributions predicted by the wave equations of quantum physics, so it is unclear whether or not such an interpretation could be a logical explanation of the apparent randomness we observe.  However, the logic of this section does prove that, fundamentally, the MUH is compatible with universes that are internally observed to have true randomness.  This explanation would also neatly explain the Feynman sum over histories approach without resorting to the confusing notion of having infinitely many histories \citep[p.180]{Greene04}.

\subsection{An Inconsistent Universe?}

One of the premises of the argument in \secref{sec_axiomatization} was that the universe must be consistent.  Although the logic already presented for asserting the consistency of reality was sound, some might wonder if this somehow conflicts with special relativity.  

In particular, there is a principle in special relativity known as the Relativity of Simultaneity that shows us that two observers witnessing a sequence of events from different relativistic frames will not agree on what occurred simultaneously.  Thus, one might ask whether or not the statement, ``Events A and B occur simultaneously'' is true or false, and get a different answer from each of the two observers.  Does this imply that the universe violates the principle of excluded middle, is inconsistent, and hence not representable by an axiomatic system?

Simply put, it does not.  If special relativity were in any way inconsistent, it could not be used to make mathematical predictions about spacetime, and the theory would be immediately rejected based on inconsistencies with experimental data.

Although the two observers disagree on what they see, the events can still be recorded in the fabric of spacetime in a mathematically consistent way.  The apparent paradox of the question arises from the flawed Newtonian intuition that simultaneity is a fundamental aspect of the universe, rather than an individual perception.  

In other words, it is like asking, ``Is this painting beautiful?''  This is simply a subjective opinion that depends on the frame of reference of the observer.  The ability to formulate sentences about the universe that are not decidable does not violate the principle of excluded middle because the ambiguity in these sentences arises from an inability to objectively translate them into the language of the axiomatic system, rather than an inconsistency of the axiomatic system itself.

If the questions were changed to, ``Does John observe A to occur simultaneously with B?'' and ``Does John reply affirmatively to the question when asked if he thinks a painting is beautiful?'' then these questions could be translated into the language of the axiomatic system, and hence they would have definite answers.

\subsection{Lack of Experimental Evidence}

Tegmark's MUH has been criticized as being untestable and unsupported by physical evidence \citep{Ellis99}.  This is not true of the ROE, because in \secref{sec_axiomatization}, it was shown that the ROE can be logically derived out of consistency with the most basic of observations.  Thus, there are simply no further experiments necessary.

However, there is one physical prediction that \emph{might} be loosely inferred on the basis of the ROE: if the Many Worlds interpretation of quantum physics is correct, then the most \emph{parsimonious} assumption is that these worlds would be logically disjoint, and hence traveling between them via wormholes or other mechanisms would be fundamentally impossible.  In this case, they could not be described as `parallel' and the concept of a `multiverse' would be nonsensical without any shared logical framework.  This implication was already mentioned in \citet{Ellis99} and \citet{Stoeger04}, although it should be noted that this is not a logical requirement, because it is also possible to have an axiomatic system that describes an entire multiverse.

\begin{comment}
It is tempting to say that the ROE also makes the Many Worlds interpretation the most parsimonious explanation of quantum physics, because the ROE already implies the existence of infinitely many worlds.  The ROE certainly makes this interpretation more plausible, but it would be hasty to say that it becomes the most parsimonious, because it remains to be shown that the presumption of Many Worlds can actually 
\end{comment}

\subsection{Theistic Objections} \label{sec_theistic}

It may seem at first that the ROE is incompatible with the concept of a God.  It certainly is consistent with the notion of a godless reality, because it shows us that a reality can be perceived from some axiomatic system without the need for a God.  However, the recognition that self-awareness \emph{can} be created axiomatically is also compelling evidence to believe that there are some realities that do have something like a God.  This is because if self-awareness can be derived in an axiomatic system via emergent phenomena, then there must also be an axiomatic system that derives self-awareness more directly without using emergent phenomena.

The fact that humans can have thoughts that are translated into physical behaviors is proof that thoughts are part of the axiomatic system that defines reality.  According to proponents of the Copenhagen interpretation, the role of observers is even more direct evidence of this.  Thus, there is no reason why an axiomatic system could not be centrally defined about the thoughts of a single self-aware entity, such that the thoughts of that entity could influence or control the reality of that system to any arbitrary degree necessary in order to meet all our criteria for being a God.

\section{Conclusion}

We have seen that Tegmark's controversial postulate can be eliminated, and that the ROE can be derived either from the assumption that self-aware entities are represented by finite information in our universe (which is an assumption that agrees with modern theories of physics and cosmology), or merely from the fact that we exist (because it was shown that existence in an objective sense leads to contradiction, and the only alternative is relative existence, which directly implies the ROE).  

Moreover, we have seen that the ROE is the only possible theory that could ever fully answer Leibniz's question or resolve the fine-tuning problem via anthropic reasoning, and that this reasoning also explains why our universe is maximally biophilic, has simple and elegant laws of nature, as well as why it is approximately causal and local at large scales but not at small scales.  Thus, the conclusion of this paper is that we must accept that the ROE is true, or rather, that existence is relative.

In closing, the recognition that self-awareness is somehow derivable from within an axiomatic system is enough to answer the fundamental question of \emph{why} we exist, but until we truly understand the actual physics of self-awareness, we cannot begin to fathom what properties are necessary for an axiomatic system to derive self-awareness in order to answer the question of \emph{how} we exist.

\bibliographystyle{plainnat}
\bibliography{reality}

\ifx \kindle \undefined
\end{multicols}
\fi

\end{document}